\documentstyle[12pt,epsfig,color]{article}
\begin{document}

\begin{center}
{\Large\bf Unitary evolution for anisotropic quantum cosmologies: models with variable spatial curvature }
\\[15mm]
Sachin Pandey\footnote{E-mail: sp13ip016@iiserkol.ac.in}, ~~
Narayan Banerjee\footnote{E-mail: narayan@iiserkol.ac.in}

{\em Department of Physical Sciences,~~\\Indian Institute of Science Education and Research Kolkata,~~\\Mohanpur Campus, Mohanpur, West Bengal-741246, India}\\[15mm]
\end{center}

\vspace{0.5cm}
\vspace{0.5cm}
\pagestyle{myheadings}
\newcommand{\be}{\begin{equation}}
\newcommand{\ee}{\end{equation}}
\newcommand{\bea}{\begin{eqnarray}}
\newcommand{\eea}{\end{eqnarray}}

\begin{abstract}
Contrary to the general belief, there has recently been quite a few examples of unitary evolution of quantum cosmological models. The present work gives more examples, namely Bianchi type VI and type II. These examples are important as they involve varying spatial curvature unlike the most talked about homogeneous but anisotropic cosmological models like Bianchi I, V and IX. We exhibit either explicit example of the unitary solutions of the Wheeler-DeWitt equation, or at least show that a self-adjoint extension is possible.
\end{abstract}

\vskip 1.0cm



\section{Introduction}

A quantum description of the universe should emerge from a  quantum theory of gravity which still eludes the reach in a generally accepted form. Quantum cosmology is a moderately ambitious programme where quantum mechanical principles are employed in a gravitational system in the absence of a more general quantum theory of gravity.  Of course quantum cosmology has its own motivation, such as looking for a resolution of the problem of singularity at the birth of the universe. The basic framework for quantum cosmology is provided by the Wheeler-DeWitt  equation\cite{dewitt, wheeler, misner}. Amongst the infinitely many possible metric, only a particular form is normally chosen  by hand from the consideration of symmetry. This is the usual minisuperspace which reduces the degrees of freedom to a finite number and thus makes the problem tractable. There are quite a few reviews which discuss the development of the subject and some of its conceptual problems\cite{wilt, halli, nelson1}. \\

One major problem is that the quantization of anisotropic models are believed to give rise to a non-unitary evolution of the wave function resulting in a nonconservation of probability. It is interesting to note that this non-unitarity is often apt to be invisible in the absence of a properly oriented scalar time parameter in the scheme of quantization\cite{lidsey, nelson2}. In a relativistic theory, time itself is a coordinate and fails to be the scalar parameter against which the evolution should be studied. In fact the problem of the proper identification of time in quantum cosmology is a subject by itself and dealt with by many\cite{kuchar1, isham, rovelli, anderson}.\\ 

A novel idea about the identification of time through the evolution of a fluid present in the model appeared to work very well. The method, where the fluid variables are endowed with dynamical degrees of freedom through some thermodynamic potentials\cite{schutz1, schutz2}, was suggested by Lapchinskii and Rubakov\cite{rubakov}. It has been shown that the time parameter that emerges out of the fluid evolution has the required monotonicity as well as the correct orientation\cite{sridip1}. This Schutz formalism is now very widely used in quantizing cosmological models\cite{sridip1, alvarenga1, alvarenga2, alvarenga3, barun, almeida, sridip2, sridip3}.\\

Until very recently, the non-conservation of probability in anisotropic models had almost been generally accepted as a pathology, and had been ascribed to the hyperbolicity of the Hamiltonian\cite{alvarenga3}. Not that the anisotropic models 
are of utmost importance so far as the observed universe is concerned, but this feature of non-unitarity renders the quantization scheme vulnerable. Also, the formation of structure in the universe indeed requires a small but finite anisotropy of  $\frac{\Delta \rho}{\rho} \sim 10^{-5}$. \\

There has now been a new turn in this picture. Majumder and Banerjee\cite{barun} showed that a suitable ordering of operators can lead to a alleviation of the problem, meaning that the probability is conserved except for a small period of time. Later it was clearly shown by Pal and Banerjee\cite{sridip1, sridip2} that the said non-unitarity can actually be attributed to either an ordering of operators or to a bad choice of variables. With a suitable ordering, examples of unitary evolution were exhibited in Bianchi I, V and IX models. The degree of difficulty in integration allowed only a few cases of choice of $\alpha$ which determines the equation of state ($ P = \alpha \rho$) for which the desired unitarity was established. However, even a few examples are good enough to indicate that the problem is not actually pathological and can be cured. Very recently an example of a unitary evolution for a Kanotowki-Sachs model has been given by Pal and Banerjee\cite{sridip3}. It was also shown by Pal\cite{sridip4} that this unitarity is achieved not at the cost of anisotropy itself. \\

Except for the Kantowski-Sachs cosmology, all other examples of the anisotropic Bianchi models stated have one unifying feature, they are all of constant spatial curvature. The motivation for the present work is to show that the possiblity of a  self adjoint extension and hence a unitary evolution is not a characteristic of models with a constant spatial curvature, this is in fact more general and can be extended to models with variable spatial hypersurfaces as well. Two specific examples, namely Bianchi II and VI are dealt with in the following sections. Section 2 deals with the formalism and takes up the example of the Bianchi VI model. Section 3 deals with the Bianchi II model. The last section includes a summary and a discussion of the results obtained.   \\

\section{The formalism and Bianchi VI models}

We start with the standard Einstein-Hilbert action for gravity along with a perfect fluid given by 

\begin{equation}
\label{action}
{\mathcal A} = \int_M d^4x\sqrt{-g}R +\int_M d^4x\sqrt{-g}P,
\end{equation}

where $R$ is the Ricci Scalar, $g$ is the determinant of the metric and $P$ is the pressure of the ideal fluid. The first  term corresponds to the gravity sector and the second term is due to the matter sector. Here we have ignored the contributions from boundary as it would not contribute to the variation. The units are so chosen that $16\pi G =1$.\\

A Bianchi VI model is given by the metric
\begin{eqnarray}
ds^2 = n^2(t)dt^2-a^2(t)dx^2-e^{-mx}b^2(t)dy^2-e^xc^2(t)dz^2,
\label{metric-6}
\end{eqnarray}

where the lapse function $n$ and $a, b, c$ are functions of time $t$ and $m$ is a constant. \\

From the metric given above, we can write the Ricci Scalar as
\begin{eqnarray}
\label{ricci-6}
\sqrt{-g}R= e^{\frac{(1-m)x}{2}} \bigg[\frac{d}{dt}[\frac{2}{n}(\dot{a}bc +\dot{b}ca+a\dot{c}b)] -\frac{2}{n}[\dot{a}\dot{b}c +\dot{b}\dot{c}a+\dot{c}\dot{a}b+\frac{n^2bc}{4a}(m^2-m+1)]\bigg].
\end{eqnarray}

Using this, we can find the  action for the gravity sector from  equation (\ref{action}) which is given as
\begin{equation}
\label{action-grav}
{\mathcal A}_g=\int dt \bigg[-\frac{2}{n}[\dot{a}\dot{b}c+\dot{b}\dot{c}a+\dot{c}\dot{a}b+\frac{n^2bc}{4a}(m^2-m+1)]\bigg],
\end{equation}
where an overhead dot indicates a derivative with respect to time. \\

Now  we make a set of transformation of variables as 
\begin{eqnarray}
a(t)=e^{\beta_0}, \\ 
b(t)=e^{\beta_0+\sqrt{3}(\beta_+-\beta_-)}, \\
c(t)=e^{\beta_0-\sqrt{3}(\beta_+-\beta_-)}.
\end{eqnarray}
This introduces a constraint $a^2=bc$, but the model is still remains Bianchi Type VI without any loss of the typical characteristics of the model. Such type of transformation of variables has been extensively used in the literature\cite{sridip1, barun, alvarenga3}. One can now write the Lagrangian density of the gravity sector as 
\begin{equation}
{\mathcal L}_g = -6\frac{e^{3\beta_0}}{n}[\dot{\beta_0^2}-(\dot{\beta_+}-\dot{\beta_-})^2 +\frac{e^{-2\beta_0}n^2(m^2-m+1)}{12}]. \label{7}
\end{equation}

Here $\beta_0$ ,$\beta_+$ and $\beta_-$ has been treated as coordinates. So corresponding Canonical momentum will be $p_0$, $p_+$ and $p_-$ where $p_{i} = \frac{\partial {\mathcal L}_g}{\partial \dot{\beta_{i}}}$. It is easy to check that one has $p_+ =-p_-$. Hence we can write the corresponding Hamiltonian as
\begin{equation}
{\mathcal H}_g=-n e^{-3\beta_0}[\frac{1}{24}(p_0^2-p_+^2-12(m^2-m+1)e^{4\beta_0})]. \label{8}
\end{equation}

With the widely used technique, developed by Lapchinskii and Rubakov\cite{rubakov} by using the Schutz formalism of writing the fluid parameters in terms of thermodynamic variables\cite{schutz1, schutz2}, the action the fluid sector can be written as
\begin{equation}
\label{action-matter}
{\mathcal A}_f =\int dt {\mathcal L}_{f}\\ = \int dt \left[n^{-\frac{1}{\alpha}}e^{3\beta_{0}}\frac{\alpha}{\left(1+\alpha\right)^{1+\frac{1}{\alpha}}}\left(\dot{\epsilon}+\theta\dot{S}\right)^{1+\frac{1}{\alpha}}e^{-\frac{S}{\alpha}}\right].
\end{equation}

Here $\epsilon, \theta, S$ are thermodynamic potentials. A constant volume factor $V$ comes out of the integral in both of (\ref{action-grav}) and (\ref{action-matter}). This $V$ is inconsequential as it can be absorbed in the subsequent variational principle. With a canonically transformed set of variables $T,\epsilon^{\prime}$ in place of $S, \epsilon$, one can finally write down the Hamiltonian for the fluid sector as

\begin{equation}
{H}_f = n  e^{-3\beta_0}e^{3(1-\alpha)\beta_0}p_T.
\label{43}
\end{equation}

The canonical transformation is given by the set of equations

\begin{eqnarray}\label{canonical}
T&=&-p_{S}\exp(-S)p_{\epsilon}^{-\alpha -1},\\
p_{T}&=&p_{\epsilon}^{\alpha+1}\exp(S),\\
\epsilon^{\prime}&=&\epsilon+\left(\alpha+1\right)\frac{p_{S}}{p_{\epsilon}},\\
p_{\epsilon}^{\prime}&=&p_{\epsilon},
\end{eqnarray}

This method and the canonical nature of the transformation are comprehensively discussed in reference \cite{sridip1}. \\

The net or the super Hamiltonian is
\begin{equation}
H= H_g + H_f = -\frac{ne^{-3\beta_0}}{24}[p_0^2-p_+^2-12(m^2-m+1)e^{4\beta_0}-e^{3(1-\alpha)\beta_0}p_T] .
\label{44}
\end{equation}
Using the Hamiltonian constraint $H=0$, which can be obtained by varying the action ${\mathcal A}_{g}+ {\mathcal A}_{f}$ with respect to the lapse function $n$, one  can write the Wheeler-DeWitt equation as
\begin{equation}
[e^{3(\alpha-1)\beta_0}\frac{\partial^2}{\partial \beta_0^2}-e^{3(\alpha-1)\beta_0}\frac{\partial^2}{\partial \beta_+^2}+12(m^2 - m + 1)e^{(3\alpha+1)\beta_0}]\psi =24i\frac{\partial}{\partial T}\psi.
\label{45}
\end{equation}
This equation is obtained after we promote the momenta to the corresponding operators given by $p_{i}=-i\frac{\partial}{\partial {\beta}_{i}}$ in the units of $\bar{h}=1$. \\

It is interesting to note that for a particular value of $m=m_0$ where $m_0$ is a root of equation $m^2-m+1 = 0$, the spatial curvature vanishes and the equation (\ref{45}) reduces to the corresponding equation for a Bianchi Type I model\cite{sridip1}. We shall discuss the solution of the Wheeler-DeWitt equation in two different cases, namely $\alpha = 1$ and $\alpha \neq 1$. \\ 

\subsection{Stiff fluid: $\alpha = 1$}

For a stiff fluid ($P=\rho$), the equation (\ref{45}) becomes simple and easily separable. It looks like
\begin{equation}
\bigg[\frac{\partial^2}{\partial \beta_0^2}-\frac{\partial^2}{\partial \beta_+^2}+12(m^2-m+1)e^{4\beta_0}\bigg]\psi  =24i\frac{\partial}{\partial T}\psi .
\label{19}
\end{equation} 

Wih the separation ansatz 

\begin{equation}
\psi = e^{i2 k_+\beta_+}\phi(\beta_0)e^{-iET},
 \label{20}
\end{equation}
one can write 
 \begin{equation}
\frac{\partial^2 \phi}{\partial \beta_0^2}+(4k_+^2-24E+4N^2 e^{4\beta_0})\phi=0,
\end{equation}
 where $N^2=3(m^2-m+1)$. After making the change in variable as $q = N e^{2\beta_0}$, above equation can be written as
 \begin{equation}
q^2\frac{\partial^2 \phi}{\partial q^2}+q\frac{\partial \phi}{\partial q}+[q^2 - (6E-k_+^2)]\phi=0.
\end{equation}
Solution of this equation can be written in terms of Bessel's functions as 
\begin{equation}
\phi(q) = J_{\nu} (q)  \label{phi_q},
\end{equation}

where $\nu =\sqrt{6E-k_+^2}$. Now for the construction of the wave packet, we need to fix $\nu$. If we take $\epsilon= -\nu^2 =k_+^2-6E$ then wave packet can have following expression 
\begin{equation}
\Psi = \Phi (q) \zeta(\beta_+) e^{i\epsilon T/6}.
\end{equation}
where \begin{equation}
\zeta(\beta_+)=\int dk_+ e^{-(k_+ -k_{+0})^2} e^{i (2k_+\beta_+ - \frac{k_+^2}{6} T)}
\end{equation}

The norm indeed comes out to be positive and finite (for the detals of the calculations, we refer to work of Pal and Banerjee \cite{sridip3}). Thus one indeed has a unitary time evolution. \\

\subsection{General perfect fluid: $\alpha \neq 1$}

Now we shall take the more complicated case of $\alpha \neq 1$ and try to solve the Wheeler-DeWitt equation (\ref{45}).
We use a specific type of operator ordering with which equation (\ref{45}) takes the form
%$$$$ 
\begin{equation}
\bigg[e^{\frac{3}{2} (\alpha-1)\beta_0}\frac{\partial}{\partial\beta_0}e^{\frac{3}{2} (\alpha-1)\beta_0}\frac{\partial}{\partial\beta_0}-e^{3(\alpha-1)\beta_0}\frac{\partial^2}{\partial\beta_+^2}
+
12(m^2-m+1)e^{(3\alpha+1)\beta_0}\bigg]\Psi=24i\frac{\partial}{\partial T}\Psi. \label{25} 
\end{equation} 

Now with the standard separation of variable as,
\begin{equation}
\Psi(\beta_0,\beta_+ ,T) =\phi(\beta_0) e^{ik_+\beta_+}e^{-iET},
\end{equation}
the equation for $\phi$ becomes
\begin{equation}
\bigg[e^{\frac{3}{2} (\alpha-1)\beta_0}\frac{\partial}{\partial\beta_0}e^{\frac{3}{2} (\alpha-1)\beta_0}\frac{\partial}{\partial\beta_0}+e^{3(\alpha-1)\beta_0}k_+^2+12(m^2-m+1)e^{(3\alpha+1)\beta_0}-24E\bigg]\phi=0. \label{26}
\end{equation}
For $\alpha \neq 1$ we make a transformation of variable as
\begin{equation}
\chi =e^{-\frac{3}{2} (\alpha-1)\beta_0},
\end{equation}
and write equation (\ref{26}) as
\begin{equation}
\frac{9}{4}(1-\alpha)^2\frac{\partial^2\phi}{\partial \chi^2}+\frac{k_+^2}{\chi^2}\phi +12(m^2-m+1)\chi^{\frac{2(3\alpha+1)}{3(1-\alpha)}}\phi-24E\phi =0. \label{27}
\end{equation}
We define some parameters as
\begin{eqnarray}
\sigma =\frac{4k_+^2}{9(1-\alpha)^2}, \\  
E' = \frac{32}{3(1-\alpha)^2}E,\\
M^2 =\frac{16(m^2-m+1)}{3(1-\alpha)^2}. \label{28}
\end{eqnarray}
Equation (\ref{27}) can now be written as
\begin{equation}
-\frac{\partial^2\phi}{\partial \chi^2}-\frac{\sigma^2}{\chi^2}\phi -M^2\chi^{\frac{2(3\alpha+1)}{3(1-\alpha)}}\phi=-E'\phi. \label{29}
\end{equation}
Above equation can be compared to $-{\mathcal H}_g=-\frac{d^2}{d\chi^2}+V(\chi)$ with $V(\chi)=-\frac{\sigma^2}{\chi^2} -M^2\chi^{\frac{2(3\alpha+1)}{3(1-\alpha)}}$ which is a continuous and real valued function on the half line, and one can show that the Hamiltonian $H_g$ admits self-adjoint extension as ${\mathcal H}_g$ has equal deficiency indices. For a systematic and detail description of the self-adjoint extension we can refer to the text by Reed and Simons\cite{reed}. \\
So it can be said that for perfect fluid with $\alpha \neq 1$ Bianchi VI quantum models do admit  a unitarity evolution.\\

\subsection{$\alpha=-\frac{1}{3}$}
We take a specific choice, where $\rho+3P =0$, as an example. This equation of state will make equation (\ref{29}) much simpler. 
With $\alpha=-1/3$, the term $-M^2\chi^{\frac{2(3\alpha+1)}{3(1-\alpha)}}$ becomes a constant ($M^{2}$). Equation (\ref{29}) becomes

\begin{equation}
-\frac{\partial^2\phi}{\partial \chi^2}-\frac{\sigma^2}{\chi^2}\phi =-(E'-M^2)\phi, \label{30}
\end{equation}
which is in fact a well known Schrodinger equation of a particle with mass $m=1/2$ in an attractive inverse square potential. Solution to above can be given as,
\begin{eqnarray}
\phi_a(\chi)=\sqrt{\chi}[AH_{i\beta}^{(2)}(\lambda \chi)+BH_{i\beta}^{(1)}(\lambda \chi)], \\
\phi_b(\chi)=\sqrt{\chi}[AH_{\alpha}^{(2)}(\lambda \chi)+BH_{\alpha}^{(1)}(\lambda \chi)], \label{31}
\end{eqnarray}
for  $\sigma >1/4$ and $\sigma < 1/4$ and $\beta = \sqrt{\sigma-1/4}$ and $\beta = \sqrt{1/4-\sigma}$ respectively. Here both $\alpha$ and $\beta$ are real numbers and in both cases the energy spectra is given as
\begin{equation}
 E'=M^2-\lambda^2.
\end{equation}

Self-adjoint extension guarantees that $|B/A|$ takes a value so as to conserve probability and make the model unitarity. The details of the calculations are omitted, as the analysis is similar to that described in reference \cite{sridip2}.

\section{Bianchi II models:}

Bianchi Type II model is given the line element
\begin{equation}
ds^2=dt^2-a^2(t)dr^2-b^2(t)d\theta^2-[a^2(t) \theta^2 +b^2(t)]d\phi^2+2a^2(t)\theta dr d\phi,
\label{51} 
\end{equation}

and the process is a bit more involved for the presence of the non-diagonal terms in the metric. \\

The Ricci scalar $R$ in this case is given by
\begin{equation}
R = -\frac{a^2}{2 b^4} - \frac{4\dot{a}\dot{b}}{a b} -\frac{2{\dot{b}}^2}{b^2} -\frac{2\ddot{a}}{a} -\frac{4\ddot{b}}{b}.
\end{equation}

If we define a new variable $\beta=a b$ as prescribed in \cite{alvarenga3}, then Lagrangian density for gravity sector looks like
\begin{eqnarray}
 {\mathcal L}_g =\frac{2\beta^2\dot{a}^2}{a^3}-\frac{2\dot{\beta}^2}{a}-\frac{a^5}{2\beta^2}, 
\label{52} 
\end{eqnarray}
 and the corresponding Hamiltonian density for gravity sector can be written as
 \begin{equation}
H_g=\frac{a^3p_a^2}{8\beta^2}-\frac{a}{8}p_{\beta}^2+\frac{a^5}{2\beta^2}. 
\label{53}
\end{equation}
Using Schutz's formalism and proper identification of time as we did before, the Hamiltonian density for fluid sector can be written as 
\begin{equation}
H_f =  a^{\alpha}\beta^{-2\alpha}p_T.
\label{54}
\end{equation}
The super Hamiltonian can now be written in following form
\begin{equation}
H = H_g + H_f = \frac{a^3p_a^2}{8\beta^2}-\frac{a}{8}p_{\beta}^2+\frac{a^5}{2\beta^2} +a^{\alpha}\beta^{-2\alpha}p_T .
\label{55}
\end{equation}

As an example we take up the case of a stiff fluid given by $\alpha=1$. \\

After promoting the momenta by operators as usual, the Wheeler-DeWitt equation  $H\Psi=0$ takes following form
\begin{equation}
-\frac{a^2}{8}\frac{\partial^2 \psi}{\partial a^2}+\frac{\beta^2}{8}\frac{\partial^2 \Psi}{\partial \beta^2}+\frac{a^4}{2}\Psi = i \frac{\partial \Psi}{\partial T}.
\label{57}
\end{equation}
Using a separation of variables
\begin{equation}
\Psi =e^{-iET}\phi(a)\psi(\beta),
\end{equation}

we get following equations for $\psi$ and $\phi$ respectively
\begin{eqnarray}
-\frac{d^2\psi}{d\beta^2}+\frac{8k}{\beta^2}\psi=0, \label{58}\\
a^2\frac{d^2\phi}{da^2}-4a^4\phi-8(k-E)\phi=0. \label{59}
\end{eqnarray}

With $\phi=\frac{\phi_0}{\sqrt{a}}$ and $\chi=a^2$, last equation can be written as
\begin{equation}
-\frac{d^2\phi_0}{d\chi^2}-\frac{\sigma}{\chi^2}\phi_0=-\phi_0 ,\label{60}
\end{equation}
where $\sigma=[\frac{3}{16}-2(k-E)].$
\\
Now equations (\ref{58}) and (\ref{60}) are governing equations for Bianchi Type II with a stiff fluid. \\

Equations for both $\psi$ and $\phi$ can be mapped to a Schrodinger equation for a particle in an inverse square potential. In order to get a solution we actually have ensure an attractive regime, which requires $k\leq 0$ ,  $E \leq k-3/32$. We see that both the equations are that for inverse square potentials, and thus a self-adjoint extension is possible. This case is actually very similar to the Bianchi IX model as discussed in refefernce \cite{sridip2}. So we do not discuss this in detail.

\section{Discussion and conclusion}
The present work deals with two examples of anisotropic quantum cosmological models with varying spatial curvature, namely Bianchi VI and II. We show that there is indeed a possibility of finding unitary evolution of the system. The earlier work on anisotropic models with constant spatial curvature\cite{sridip1, sridip2} disproved the belief that anisotropic quantum cosmologies generically suffer from a pathology of non-unitarity.  The present work now strongly drives home the fact that this feature is not at all a charactristic of models with constant spatial curvature. It was also shown before that the unitarity is not achieved at the cost of anisotropy itself\cite{sridip4}. One can now indeed work with quantum cosmologies far more confidently, as there is actually no built-in generic non-conservation of probability in the models. \\

Very recently it has been shown that in fact all homogeneous models, isotropic or anisotropic, quite generally have a self-adjoint extension\cite{sridip5}. The present work gives two more examples, and consolidates the result proved in reference \cite{sridip5}. The extension, however, is non-unique in anisotropic models. \\

Thus the standard canincal quantization of cosmological models via Wheeler-DeWitt equation still proves to be useful in the absence of a more general quantum theory of gravity. The more challenging work will now be the quantization of inhomogeneous cosmological models.

\vskip 1.50cm

{\bf Acknowledgment} The authors thank Sridip Pal for stimulating discussions. SP thanks the CSIR (India) for financial support.

\end{document}